\documentclass[preprint,showpacs,amsmath,amssymb,showkeys]{revtex4}
\usepackage{amssymb}
\usepackage{amsmath}
\usepackage{graphicx}
\usepackage{epsfig}

\begin{document}

\title{Geometrical Properties of Coupled Oscillators at Synchronization}
\author{Hassan F. El-Nashar{\footnote{
Electronic address :hfelnashar@gmail.com}}}
\affiliation{Department of Physics, Faculty of Science, Ain Shams University, 11566
Cairo, Egypt}
\affiliation{Department of Physics, Faculty of Science and Humanitarian Studies, Alkharj University, P.O.~Box
21034, 11942 Alkharj, K.S.A}
\author{Hilda A. Cerdeira{\footnote{
Electronic address(corresponding author): cerdeira@ift.unesp.br}}}
\affiliation{Instituto de F\'{\i}sica Te\'orica, UNESP-Universidade Estadual Paulista,
Caixa Postal 70532-2, 01156-970, S\~ao Paulo, SP, Brazil}
\date{\today }

\begin{abstract}
We study the synchronization of $N$ nearest neighbors coupled oscillators in
a ring. We derive an analytic form for the phase difference among
neighboring oscillators which shows the dependency on the periodic boundary
conditions. At synchronization, we find two distinct quantities which
characterize four of the oscillators, two pairs of nearest neighbors, which
are at the border of the clusters before total synchronization occurs. These
oscillators are responsible for the saddle node bifurcation, of which only
two of them have a phase-lock of phase difference equals $\pm$$\pi$/2. Using
these properties we build a technique based on geometric properties and
numerical observations to arrive to an exact analytic expression for the
coupling strength at full synchronization and determine the two oscillators
that have a phase-lock condition of $\pm$$\pi$/2.
\end{abstract}

\maketitle

\section{Introduction}

Coupled oscillators have been used to understand the behavior of
systems in physics, chemistry, biology, neurology as well as other
disciplines. In particular, they are used to model phenomena such as:
Josephson junction arrays, multimode lasers, vortex dynamics in fluids,
biological information processes, neurodynamics \cite{1,2,3}. These systems
have been observed to synchronize themselves to a common frequency, when the
coupling strength between the oscillators is increased \cite{3,4,5}.
Although all these phenomena have different dynamics, their synchronization
features might be described using a simple model of weakly coupled phase
oscillators such as the Kuramoto model and variations to adapt it for finite
range interactions which are more realistic to represent physical systems
\cite{3,4,5,6}. But, finite range interactions difficult the analysis and
search for analytical solutions. In spite of that, in order to figure out
the collective phenomena when finite range interactions are considered, it
is of fundamental importance to study and to understand the case of nearest
neighbor interactions, which is the simplest form of the local interactions.
In this context, a simplified version of the Kuramoto model with nearest
neighbor coupling in a ring topology, which we shall refer to as locally
coupled Kuramoto model (LCKM), is a good candidate to describe the behavior
of coupled systems with local interactions. The LCKM has been used to
represent the dynamics of a variety of systems. Specifically, it has been
shown that a ladder array of Josephson junctions can be expressed by a LCKM
\cite{7}. Phase synchronization, in nearest neighbors coupled R\"{o}ssler
oscillators and locally coupled lasers where local interactions are
dominant, can also be described by the LCKM \cite{8,9,10,11}. Other examples
are: the occurrence of travelling waves in neurons, chains in disorders,
multi cellular systems in biology, the dynamics of an edge dislocation in a
2D lattice and an antenna array in communication systems \cite{3,4,6,12,13,14}.

While in the Kuramoto model of long range interactions one has to get a
solution in a mean field approximation, in the local model it is necessary
to study the behavior of individual oscillators in order to understand the
collective dynamics. Due to the difficulty in applying standard techniques
of statistical mechanics in order to obtain a close picture of the effect of
the local interactions on synchronization, we rely on a simple approach to
understand the coupled system with local interactions, by means of numerical
study of the temporal behavior of the individual oscillators. In this case,
numerical investigations provide a good tool to understand the mechanism of
interactions at the stage of complete synchronization which in turn helps to
get an analytic solution. Earlier studies on the LCKM show several
interesting features including tree structures with synchronized clusters,
phase slips, bursting behavior, saddle node bifurcation and so on \cite{15,16}. It has also been shown that neighboring elements share dominating
frequencies in their time spectra, and that this feature plays an important
role in the dynamics of formation of clusters in the local model \cite{17,18}; that the order parameter, which measures the evolution of the phases of
the nearest neighbor oscillators, becomes maximum at the partial
synchronization points inside the tree of synchronization \cite{19} and a
scheme has been developed based on the method of Lagrange multipliers to
estimate the critical coupling strength for complete synchronization in the
local Kuramoto model with different boundary conditions \cite{20}. In
addition, based on numerical investigations, we identified two oscillators
which are responsible for dragging the system into full synchronization \cite{21}, and the difference in phase for this pair is $\pm$$\pi$/2. These two
oscillators are among two pairs of oscillators which are formed by the four
oscillators at the borders between major clusters in the vicinity of the
critical coupling. Using these findings we developed a method to obtain a
mathematical expression for the estimated value of the critical coupling at
which full synchronization occurs, once a set of initial conditions for the
frequencies of the $N$ oscillators is assigned \cite{22}.

In this work, we use the fact that at the stage of full synchronization, all
oscillators have a common frequency and the time dependence of the phase
difference among neighboring oscillators will vanish. Using this we are able
to derive an analytic expression for the phases of oscillators, which shows
the actual dependence on the periodic boundary condition. It is clear that
this effect will decrease when the number of oscillators $N$ increases, and
the result will converge to that of a free chain as $N$ tends to infinity
\cite{22}. Even so, both problems are different and have interesting
peculiarities when the number of oscillators is finite. This problem is not
a mere detail in a theoretical problem since a ring with a finite number of
oscillators has many applications in electronics, coupled lasers and in
communications \cite{13,23,24}. In the process of finding the solution,
using the saddle node bifurcation which dominates the dynamics at
synchronization, we come across two quantities which will permit us to
identify the four oscillators at the borders between major clusters, where
only two of them will have a phase-lock condition of $|\pi/2|$. We use the
properties of these oscillators and the ones at the boundaries to define a
triangle and use its trigonometric properties to derive an analytic formula
for the critical coupling. Finally, we can identify directly the pair of
oscillators which has the phase-lock condition, which depends only on the
set of initial frequencies.

This paper is organized as follows: In Section II, we introduce the LCKM
with boundary conditions. We determine analytically the critical coupling at
the stage of complete synchronization. Finally, in Section III we give a
conclusion which is based on a summary of the results.

\section{The Model}

The LCKM can be considered as a diffusive version of the Kuramoto model, and
it is expressed as

\begin{equation}  \label{GrindEQ__1_}
\dot{\theta }_{i} =\omega _{i} +K\left(\sin \phi _{i} -\sin \phi _{i-1}
\right)
\end{equation}
with periodic boundary conditions $\theta _{i+N} =\theta _{i} $ and phase
difference $\phi _{i} =\theta _{i+1} -\theta _{i} $ for $i=1,2,...,N$. The
set of the initial values of frequencies $\omega_i$ are the natural
frequencies which are taken from a Gaussian distribution and $K$ is the
coupling strength. These nonidentical oscillators of system 1 cluster in
time averaged frequency, under the influence of the coupling, until they
completely synchronize to a common value given by the average frequency $
\omega_o =\frac{1}{N} \sum_{i=1}^{N}\omega_i$ at a critical coupling $K_c$
as shown in Figure 1. At the vicinity of $K_c$, there are only two clusters
of successive oscillators whose borders are nearest neighbors. The
oscillators remain synchronized for $K \ge K_c$ when all phase differences
and frequencies are constants. In Figure 1, we show the synchronization tree
for a system with $N=40$ oscillators, where the elements which compose each
one of the major clusters, just before complete synchronization occurs, are
indicated in each branch. These clusters merge into one at $K_c$ where all
oscillators have the same frequency. The major clusters, at the onset of
synchronization, just before $K_c$ contain $N_1$ and $N_2$ oscillators,
where $N=N_1+N_2$. It is not necessary for these clusters to have the same
numbers of oscillators. We point to the oscillators at the border of each
cluster, which are formed with successive elements, thus their bordering
elements are nearest neighbors, as mentioned above; i.e, $\ell, \ell+1, m$
and $m+1$. An interesting fact emerges here: there is a phase-locked
solution, where the phase difference between two oscillators is $\pm\pi/2$,
and it is always valid for one and only one phase difference. This can be
the difference between phases of any two nearest neighbors of the four
oscillators at the border of the clusters at the onset of synchronization
\cite{21}. However, we can not directly allocate these two oscillators
unless we do it numerically. At this point, we remind the reader that the
location of each oscillator corresponds to a well defined entity,
characterized by an initial frequency, therefore wondering about the meaning
of the location of these borders as well as the position of the boundaries,
corresponds to knowing which of the many oscillators in our systems will
dominate the dynamics.
%\begin{figure}[!ht]
%\centering
%\includegraphics[width=\linewidth,clip]{fig1.eps}
%\caption{Synchronization tree for a system of 40 oscillators with detailed
%composition of each cluster before full synchronization. Here we point the
%oscillators at the border}
%\end{figure}

Thus, the time evolution of the phase differences among neighboring
oscillators will be written as
\begin{equation}  \label{GrindEQ__2_}
\dot{\phi }_{i} =\Delta _{i} -2K\sin (\phi _{i} )+K\sin (\phi _{i-1} )+K\sin
(\phi _{i+1} ),
\end{equation}
where $\Delta _{i} =\omega _{i+1} -\omega _{i} $. We use the fact that at $
K_c$, the quantities $\dot{\phi }_{i} =0$, to derive the following
expression
\begin{equation}  \label{GrindEQ__3_}
\sin (\phi _{i} )=\frac{H_{i} }{K_{c} } +\sin (\phi _{N} ),
\end{equation}
where, $H_{i} =\frac{(N-i+1)}{N} \left[\sum _{i=1}^{N-1}i\Delta _{i}
+\left(\sum _{j=1}^{i-1}j\Delta _{j} \right)\delta _{ij} \right]$. According
to equation 3, we need to identify not only the two oscillators, say $j$ and
$j+1$, which will phase-lock with $\phi _{j} =|\pi /2|$, but also the value
of the phase difference between the oscillators identified as boundaries to find the critical coupling. In order to tackle this difficulty we
depend on numerical simulations of system 1 for different number of
oscillators and sets of initial frequencies $\omega_i$, as well as on
studying the quantity $H_i$. It should be noted that the quantity $\max
\left\{|H_{i} |\right\}$ determines the value of the critical coupling and
the oscillators locked on $|\pi/2|$ at synchronization, for the case of a
chain of free ends (open boundaries). This can be clearly understood from
equation 3, when $\sin (\phi _{N} )=0$, and $|\sin (\phi _{j} )|=1$, which
coincides with the same index of $\max \left\{|H_{i} |\right\}$, and
henceforward $K_{c} =\max \left\{|H_{i} |\right\}$. However, when periodic
boundary conditions are turned on, this fact does not hold true, since $\sin
(\phi _{N} )$ can be either positive or negative. Therefore, its presence in
equation 3 modifies the quantity $H_j$ which matches the same index of the
phase difference which locks to $|\pi/2|$, and the oscillators no longer
synchronize to a single frequency $\omega _{o} $ at a critical coupling
defined as $\max \left\{|H_{i} |\right\}$. Helped by numerical simulations,
we can find the four oscillators at the edges of the two clusters at the
onset of synchronization, in the vicinity of $K_c$, and see that only one
phase difference is locked to $|\pm$$\pi/2|$ at $K_c$ as well as to
determine which two oscillators among these four are responsible for the
saddle node bifurcation.
%
%\begin{figure}[!ht]
%\centering
%\includegraphics[width=\linewidth,clip] {fig2.eps}
%\caption{Selected values of $\sin (\protect\phi _{i} )$ at $K_c$ for the
%oscillators of Figure 1.}
%\end{figure}

When we study numerically the phase differences among neighboring
oscillators we find two distinct values among all phase differences, which
are characterized by being the absolute maximum and minimum among all phase
differences and we shall see that they correspond to the two phase
differences of the four oscillators at the borders \cite{22}. But, only one
of them will have the phase-lock condition of $|\pm\pi/2|$ at $K_c$. If we
now proceed to study the behavior of the quantity $H_i$, we see that the
maximum and minimum values of $H_i$ always coincide with the same indexes of
the maximum and minimum of $\sin (\phi _{i} )$, respectively.
%
%\begin{figure}[!ht]
%\centering
%\includegraphics[width=\linewidth,clip] {fig3.eps}
%\caption{The values of $H_i$ versus oscillator position, for the oscillators
%of Figure 1.}
%\end{figure}
%
Therefore, based on the numerical simulations, we can distinguish two
quantities $H_{\ell}$ and $H_m$ and their corresponding phase differences $%
\phi _{\ell } $ and $\phi _{m} $. Any one of them, $H_{\ell}$ or $H_m$ can
be the absolute maximum or the absolute minimum. This holds true for the two
values $\sin (\phi _{\ell } )$ and $\sin (\phi _{m} )$ also. Comparing
between the absolute values of either the two quantities $H_{\ell}$ and $H_m$
or the two quantities $\sin (\phi _{\ell } )$ and $\sin (\phi _{m} )$, we
can not say which oscillators will have the phase-lock condition. However,
we find always, and for any $N$ with any sets of $\omega_i$, that the signs
of $H_{\ell}$ and $\sin (\phi _{\ell } )$, as well as the corresponding $H_m$
and $\sin (\phi _{m} )$ are always the same, independent of being positive
or negative. Figure 2a shows the plot of selected values of $\sin (\phi _{i}
)$ versus time at $K_c$ for the same realization of Figure 1. We can
see that $\sin (\phi _{\ell } )=1$ for $\ell = 7$ and $\sin (\phi _{m} )<0$,
for $m = 23$, which is the absolute minimum among all corresponding values
of all oscillators. Figure 2b shows the values of $H_i$ versus the positions
of oscillators in the chain for the same case of Figure 1, where we clearly
see that $H_{\ell} > 0$ for $\ell = 7$ is the absolute maximum, while $H_m <
0$ for $m = 23$ is the absolute minimum.
%
%\begin{figure}[!ht]
%\centering
%\includegraphics[width=\linewidth,clip] {triangle1.eps}
%\caption{Triangle of sides: $a=(H_{\ell } )^{1/2} $, $b=(H_{\ell } -H_{m}
%)^{1/2} $ and $c=(H_{\ell } -H_{\ell -1} )^{1/2} $.}
%\end{figure}
%
Since these oscillators are obviously special, we shall focus on them in
search for the exact solution of the problem. We see that the maximum and
minimum of $H_i$ determine two major cases: $H_{\ell} > 0$ and $H_m < 0
$. Once we select the first case, we notice that, again, we encounter two
cases. We shall see that only one of them decides the exact value of the
critical coupling. These two sub-cases are a) $\sin (\phi _{\ell} )=1$ with $%
\sin (\phi _{m} )>-1$, and b) $\sin (\phi _{\ell} )<1$ with $\sin (\phi _{m}
)=-1$. For case (a) we use three equations from system \eqref{GrindEQ__3_}
for $\ell, \ell-1$ and $m$, from which we can write $\frac{\sin (\alpha /2)}{%
a} =\frac{\sin (\beta /2)}{b} =\frac{\sin (\gamma /2)}{c} $. This relation
correspond to the properties between the angles and sides of a triangle,
where the angles and sides are defined as: $\alpha =\mathrm{-}\phi _{N} +\pi
/2$, $\beta =\mathrm{-}\phi _{m} +\pi /2$ and $\gamma =\mathrm{-}\phi _{\ell
-1} +\pi /2$, $a=(H_{\ell } )^{1/2} $, $b=(H_{\ell } -H_{m} )^{1/2} $ and $%
c=(H_{\ell } -H_{\ell -1} )^{1/2} $, as shown in Figure 4. After some
manipulation, we get an expression for $\phi _{m} $ as

\begin{equation}  \label{GrindEQ__4_}
\phi _{m} =\frac{\pi }{2} -2\cos ^{-1} \left\{\frac{a^{2} +c^{2} -b^{2} }{2ac%
} \right\}.
\end{equation}
Thus, the value of the critical coupling becomes: $K_{c}^{\ell } =\frac{%
H_{\ell } -H_{m} }{1-\sin (\phi _{m} )} $. Applying the same method to case
(b) ($\sin (\phi _{l} )<1$ with $\sin (\phi _{m} )=-1$), we get $K_{c}^{m} =%
\frac{-(H_{m} -H_{\ell} )}{1+\sin (\phi _{\ell} )} $. Therefore, the value
of the critical coupling when $H_{\ell}> 0$ and $H_m < 0$ is
\begin{equation}  \label{GrindEQ__5_}
K_{c} =\max \{ K_{c}^{\ell } ,K_{c}^{m} \} .
\end{equation}
Following the same method for the case $H_{\ell} < 0$ and $H_m > 0$, we
find: $K_{c}^{\ell } =\frac{-(H_{\ell } -H_{m} )}{1+\sin (\phi _{m} )} $ and
$K_{c}^{m} =\frac{H_{m} -H_{\ell} }{1-\sin (\phi _{\ell} )} $. The value of
the critical coupling is again given by equation 5. Table 1 shows the
results from numerical simulations of system 1 which are in good agreement
with the values obtained from equation 5 for the same sets of initial
frequencies.

\begin{table}[tbp]
\caption{Values of $K_c$ from simulation of system 1 and from equation 5.}%
\begin{tabular}{|ccc|cc|cc|}
\hline
& $N$ &  & $K_c$: simulation &  & $K_c$: equation \eqref{GrindEQ__5_} &  \\
\hline
& 30 &  & 3.73094125 &  & 3.72862539 &  \\
& 40 &  & 4.29473534 &  & 4.28553971 &  \\
& 50 &  & 4.48415639 &  & 4.47935214 &  \\
& 100 &  & 5.86827841 &  & 5.86639415 &  \\
& 200 &  & 7.96802973 &  & 7.96457428 &  \\ \hline
\end{tabular}%
\end{table}

Equation 5 allows us to determine whether $\phi _{\ell } $ or $\phi _{m} $
has the phase-lock condition $|\pi/2|$ at $K_c$, depending whether the
selected value for $K_c$ is $K_{c}^{\ell } $ or $K_{c}^{m} $, respectively.
After this is done we can determine the value of $\sin (\phi _{N} )\ne 0$,
thus showing how the phase difference $\phi _{N} $, consequence of the
periodic boundary conditions, influences the dynamics of the system. We
notice from Figure 2a that the value of $\sin (\phi _{40} )\ne 0$, and it
modifies the result of the critical coupling, as expected from equation 3.
The numerical simulations, for different $N$ and for different sets of $
\omega_i$, show that the system with periodic boundary conditions has a
critical value which depends on $N$, and that the influence of $\sin (\phi
_{N} )$ decreases as $N$ increases which may give indications that it
becomes ineffective, but in this case the system never synchronizes at the
critical strength given only by the absolute maximum of $H_i$, as it should
be for a free chain. We find also, that $\sin (\phi _{N} )$ remains
different from zero up to $N \sim 2000$. Even though, $\sin (\phi _{N} )$
has a small value, but since it exists, it enforces the oscillators to
synchronize approximately as an average of $H_{\ell}$ and $H_m$, which are
not in generally equal to each other. For larger values of $N$, we expect
that $|H_{\ell}|\approx|H_m|$ and $\sin (\phi _{N} )=0$, and hence the
critical coupling is determined by the absolute value of $H_{\ell}$. As $%
N\rightarrow\infty$, the system of oscillators in a ring synchronizes at the
same value of the coupling strength as a chain of free ends, as expected
\cite{15}.

Qualitatively, we can compare the cases of synchronization for a ring and
for a free chain for finite number of oscillators in order to understand
why the critical coupling depends on both values $\max \left\{H_{i} \right\}$
and $\min \left\{H_{i} \right\}$ for the first case while the free chain
only depends on $\max \left\{|H_{i} |\right\}$. If we consider both systems
above full synchronization and decrease the coupling strength we observe
that a saddle node bifurcation occurs at the critical value and they split
into two clusters of unequally number of oscillators in general. The chain
splits into two clusters as shown in figure 5a, where only two oscillators
seem to have a new role in the dynamics, $j$ and $j+1$ are located in
different clusters. This splitting occurs when the phase difference $\phi
_{j} =|\pi /2|$, and it is the index that maximizes $\max \left\{|H_{i}
|\right\}$ which determines the value of $K_c$ for the chain. The ring
at the critical coupling splits into two clusters of unequal numbers
of oscillators, as shown in
figure 5b. The difference is
that for the ring two sets of oscillators acquire a special role in the
dynamics: $\ell$, $\ell+1$, $m$ and $m+1$ but only one phase difference will
be $\phi _{j} =|\pi /2|$ and it can be either $\phi _{\ell }$ or $\phi _{m}
$, where the two indexes $\ell$ and $m$ match $\max \left\{H_{i} \right\}$
and $\min \left\{H_{i} \right\}$, respectively. A priori we do not know
which will be. From the previous discussion we can write $K_{c} =\left|H_{j}
\pm \delta \right|_{\max }$, where $j=\ell \mathrm{\; or\; }m$ and $\delta $
depends on the periodic boundary conditions, in order to get $|\sin (\phi
_{j} )|=\left|X_{j} +\sin (\phi _{N} )\right|=1$ and $X_{j} =H_{j} /K_{c}
\ne \pm 1$ and hence $\sin (\phi _{N} )$ ``decides'' which will be the pair
of oscillators which phases-lock at $|\pi /2|$. Therefore, for finite $N$,
the oscillators synchronize at a critical coupling that depends on both
quantities $\max \left\{H_{i} \right\}$ and $\min \left\{H_{i} \right\}$,
and this is due to the presence of the periodic boundary conditions. From
equations 3 and 5 we can write: $K_{c} =\frac{|H_{\ell } |+|H_{m} |}{1+|\sin
(\phi _{j} )|} $, with $|\sin (\phi _{j} )|<1$, where $j=m$ and $|\sin (\phi
_{\ell } )|=1$, and $j=\ell$ when $|\sin (\phi _{m} )|=1$.

%\begin{figure}[!ht]
%\centering
%\includegraphics[width=\linewidth,clip] {chain1.eps}\newline
%Fig. 5(a) \newline
%\includegraphics[width=\linewidth,clip] {ring1.eps}\newline
%Fig. 5(b) \newline
%\caption{The two major clusters at the critical coupling in (a) the chain of
%free ends and (b) the ring.}
%\end{figure}

\section{Conclusions}

We have studied the synchronization of coupled oscillators in a locally
coupled Kuramoto model with nearest neighbors coupling with periodic
boundary conditions. Particularly, we analyzed the system of oscillators at
the stage of complete synchronization. We see that the dynamics is
determined by a particular set of oscillators which are located at the
borders of the major clusters which will meet at the critical coupling to
form one cluster of all synchronized oscillators. Using the boundary
conditions we can make a correspondence of the values of some definite
quantities $H_{i}$\textit{%
%TCIMACRO{\U{b4}}%
%BeginExpansion
\'{}%
%EndExpansion
s} with the sides of a triangle with the aid of some trigonometric
properties. These quantities correspond to those of the border oscillators
and their nearest neighbors. Using these two quantities, we derive a
mathematical expression for the critical coupling at synchronization. In
addition, we are also able to determine the two oscillators which will have
a phase-lock condition of $|\pi /2|$. All these properties can be calculated
a priori since they depend only on the set of initial frequencies $\omega
_{i}$.

\section*{Acknowledgement}

H F E thanks the Instituto de F\'{\i}sica T\'{e}orica, UNESP-Universidade
Estadual Paulista, S\~ao Paulo, SP, Brazil, for the hospitality during part
of this work. HAC  acknowledges support from the Conselho Nacional de Desenvolvimento Cient\'{\i}fico e Tecnol\'{o}gico (CNPq), Project
CNPq-DST.

\newpage

\begin{figure}[!ht]
\centering
\includegraphics[width=\linewidth,clip]{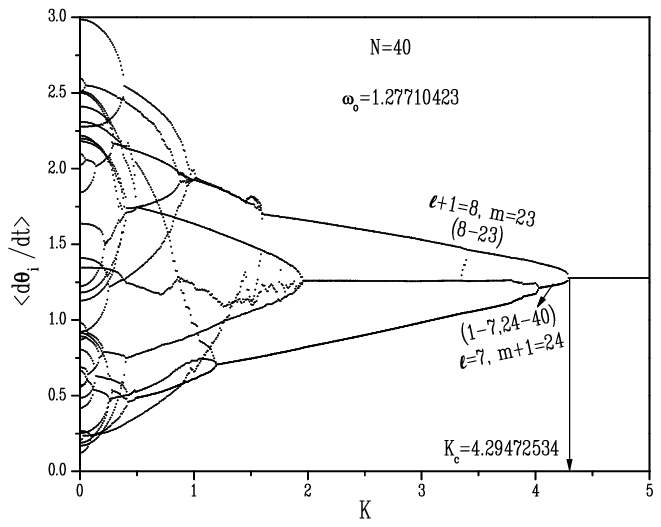}
\caption{Synchronization tree for a system of 40 oscillators with detailed
composition of each cluster before full synchronization. Here we point the
oscillators at the border}
\end{figure}

\newpage

\begin{figure}[!ht]
\centering
\includegraphics[width=\linewidth,clip] {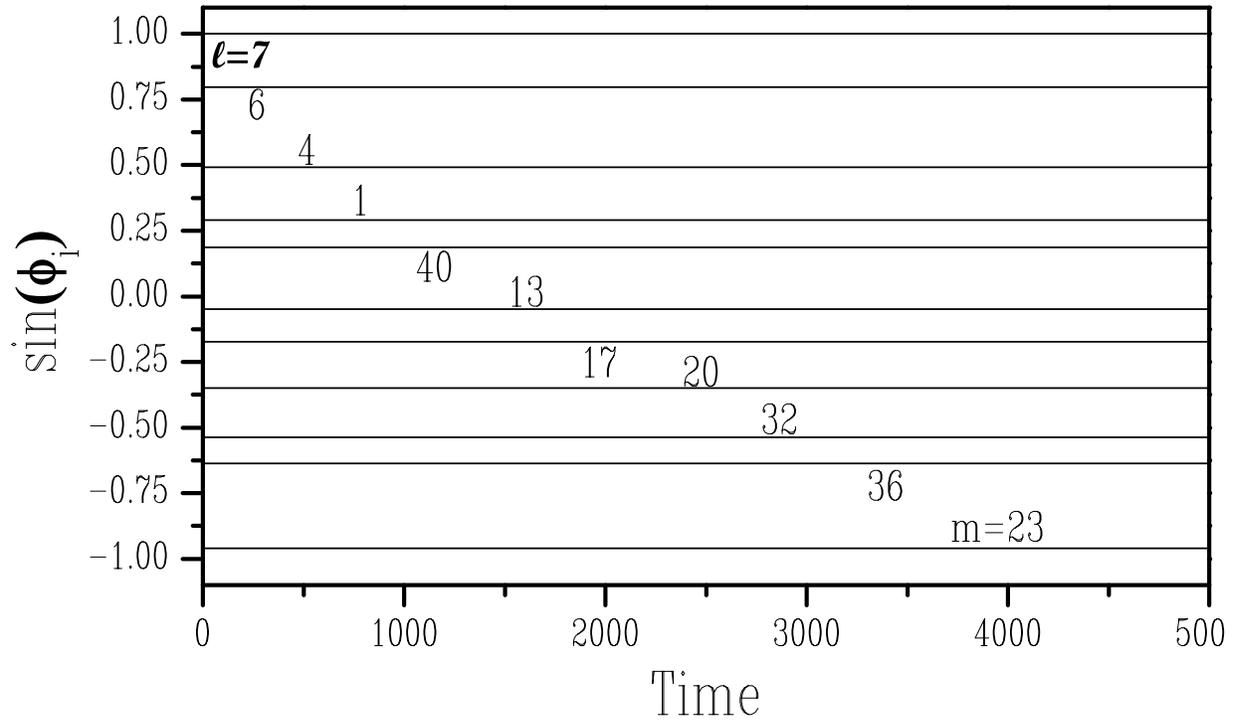}
\caption{Selected values of $\sin (\protect\phi _{i} )$ at $K_c$ for the
oscillators of Figure 1.}
\end{figure}

\newpage

\begin{figure}[!ht]
\centering
\includegraphics[width=\linewidth,clip] {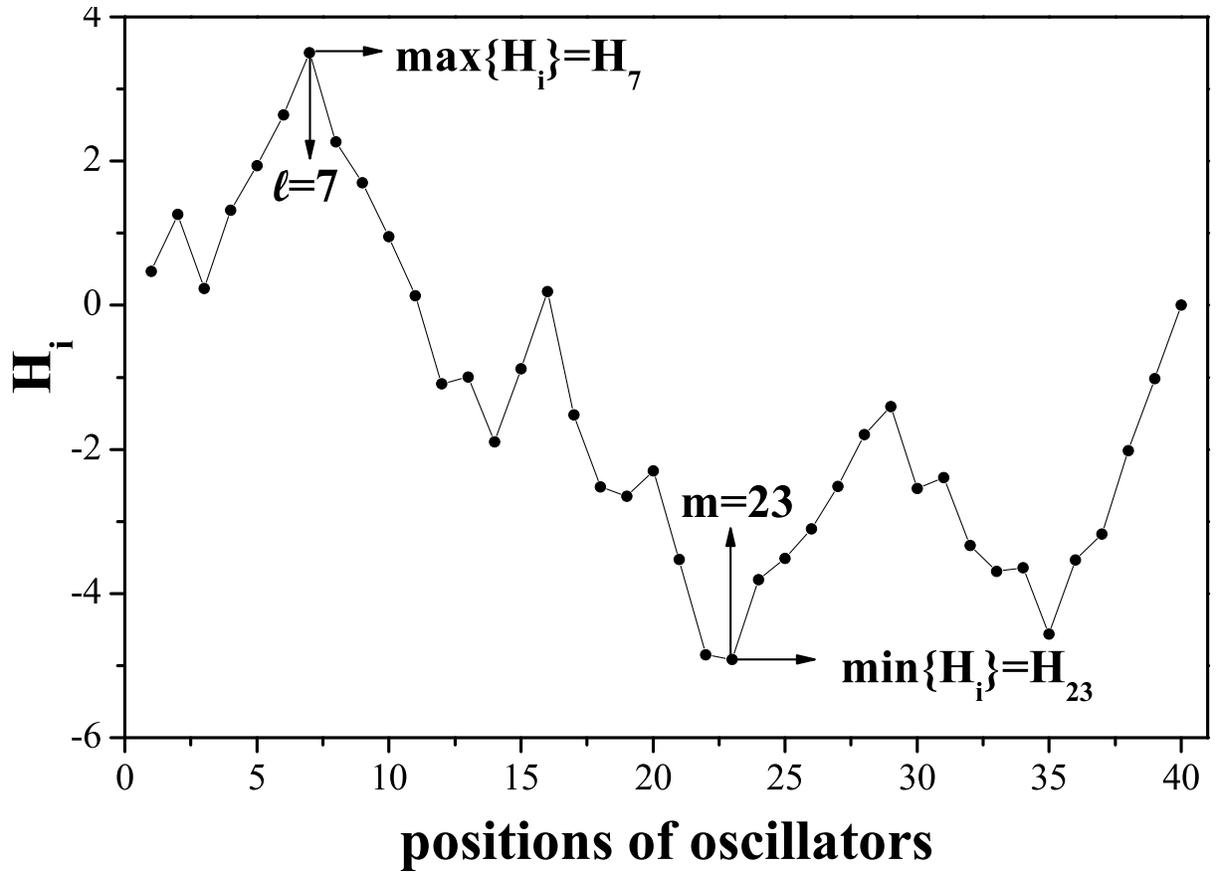}
\caption{The values of $H_i$ versus oscillator position, for the oscillators
of Figure 1.}
\end{figure}

\newpage

\begin{figure}[!ht]
\centering
\includegraphics[width=\linewidth,clip] {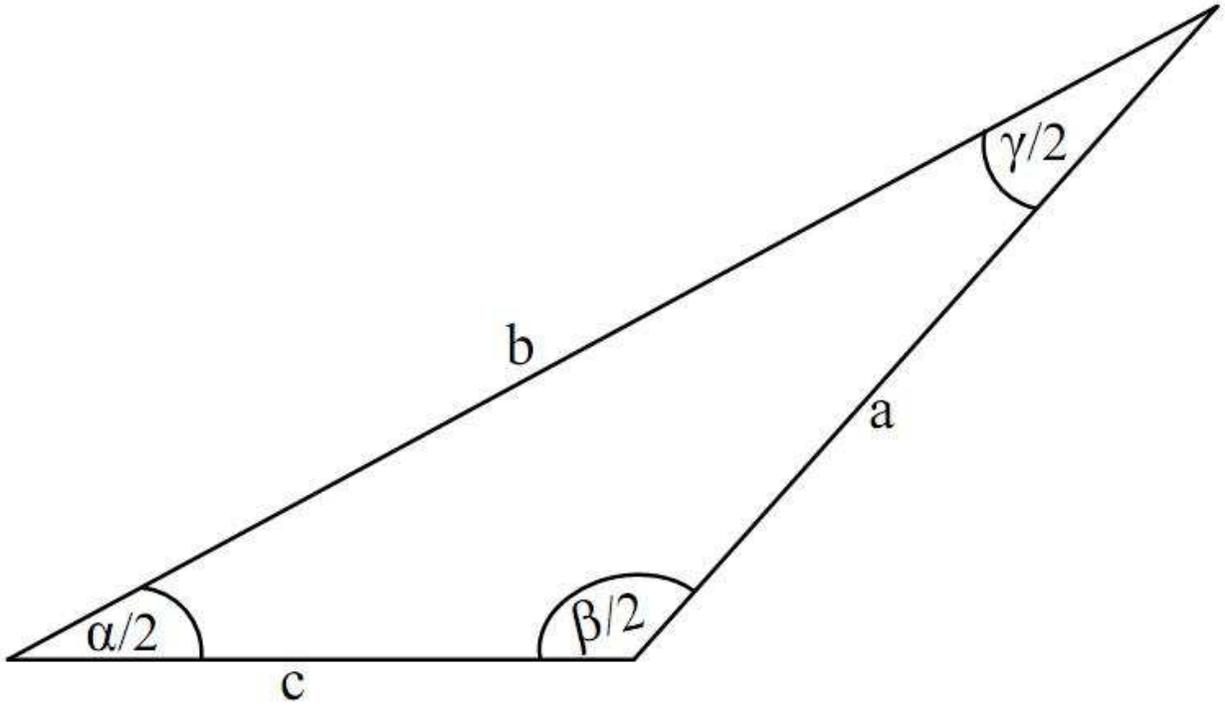}
\caption{Triangle of sides: $a=(H_{\ell } )^{1/2} $, $b=(H_{\ell } -H_{m}
)^{1/2} $ and $c=(H_{\ell } -H_{\ell -1} )^{1/2} $.}
\end{figure}

\newpage
\begin{figure}[!ht]
\centering
\includegraphics[width=\linewidth,clip] {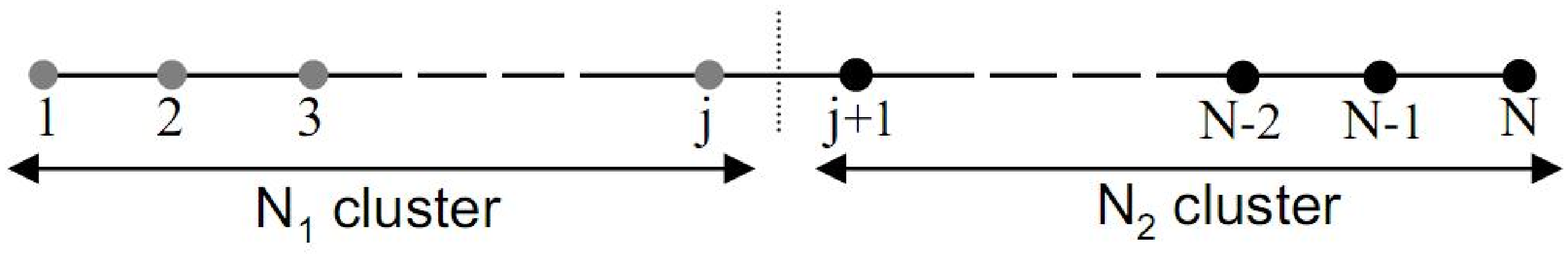}\newline
Fig. 5(a) \newline
\includegraphics[width=\linewidth,clip] {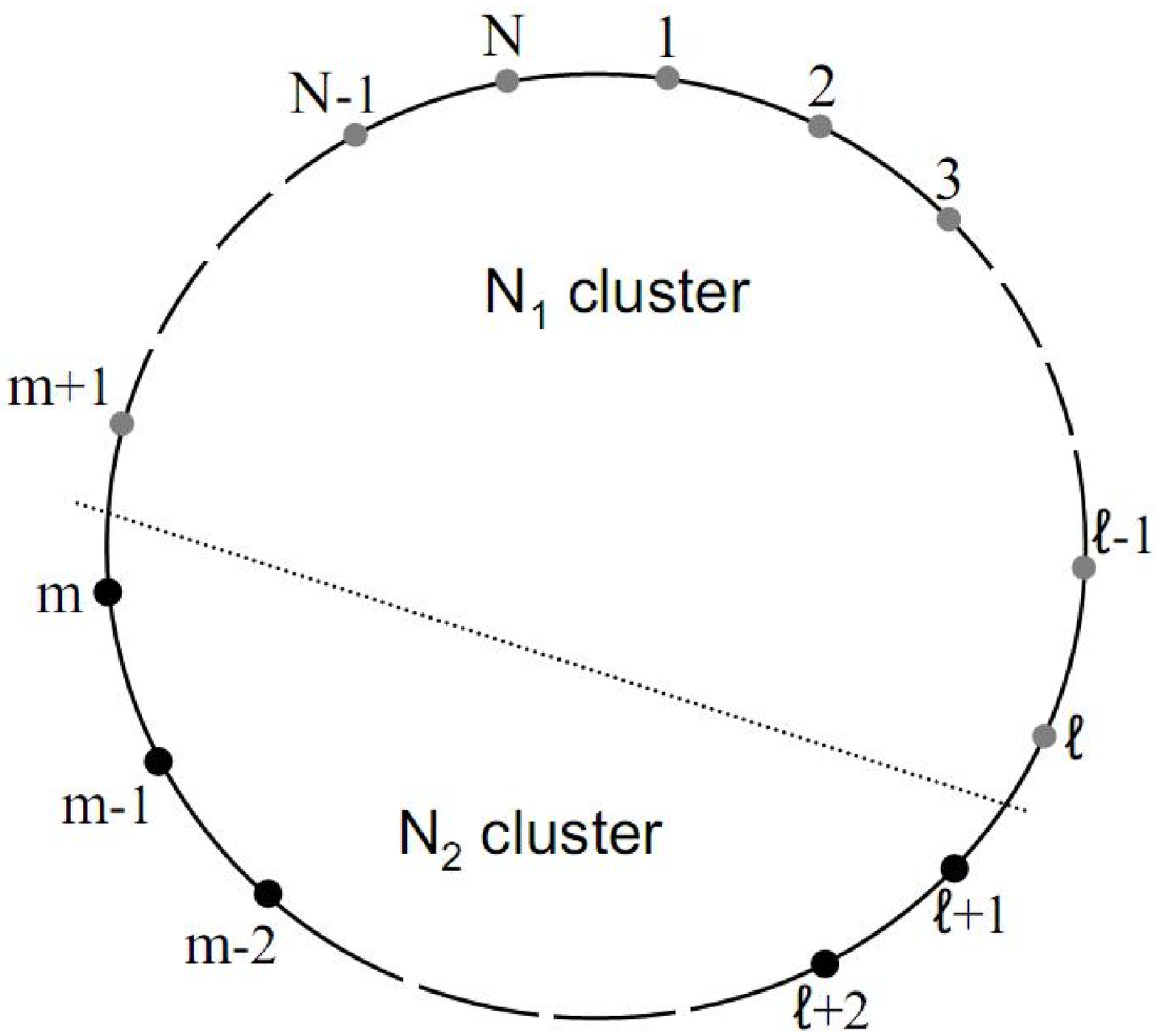}\newline
Fig. 5(b) \newline
\caption{The two major clusters at the critical coupling in (a) the chain of
free ends and (b) the ring.}
\end{figure}


\begin{thebibliography}{99}
\bibitem{1} A.T. Winfree, Geometry of Biological Time, Springer, New York, 1990.
1990.

\bibitem{2} C.W. Wu, Synchronization in Coupled Chaotic Circuits and Systems, World Scientific, Singapore, 2002.

\bibitem{3} S. Manrubbia, A. Mikhailov, D. Zanette, Emergence of Dynamical Order: Synchronization Phenomena in Complex Systems, World Scientific, Singapore, 2004.


\bibitem{4} H. Haken, Brain Dynamics: Synchronization and Activity Patterns in Pulse-Coupled Neural Nets with Delays and Noise, Springer, New York, 2007.

\bibitem{5} Y. Kuramoto, Chemical Oscillations, Waves and Turbulences, Springer, New York, 1984.

\bibitem{6} J.A. Acebron, L.L. Bonilla, C.P.J. Vicente, F. Ritort, R. Spigler, A simple paradigm for synchronization phenomena: The Kuramoto model, Rev. Mod. Phys. 77 (2005) 137-185.

\bibitem{7} B.C. Daniels, S.T.M. Dissanayake, B.R. Trees, Synchronization of coupled rotators: Josephson junction ladders and the locally coupled Kuramoto model, Phys. Rev. E 67 (2003) 026216.

\bibitem{8} Z. Liu, T.-C. Lai, F.C. Hoppensteadt, Phase clustering and transition to phase synchronization in a large number of coupled nonlinear oscillators, Phys. Rev. E 63 (2001) 055201R.

\bibitem{9} Y. Braiman, T.A. Kennedy, K. Wiesenfeld, A. Khibnik, Entrainment of solid state laser arrays, Phys. Rev. A 52 (1995) 1500-1506.

\bibitem{10} A. Khibnik, Y. Braiman, V. Protopopescu, T.A. Kennedy, K. Wiesenfeld, Amplitude dropout in coupled lasers, Phys. Rev. A 62 (2000) 063815.

\bibitem{11} D. Tsygankov D, Wiesenfeld K, Weak-link synchronization, Phys. Rev. E 73 (2006) 026222.

\bibitem{12} Y. Ma, K. Yoshikawa, Self-sustained collective oscillation generated in an array of nonoscillatory cells, Phys. Rev. E 79 (2009) 046217.

\bibitem{13} J. Rogge, D. Aeyels, Stability of phase locking in a ring of unidirectionally
coupled oscillators, J. Phys. A 37 (2004) 11135-11148.

\bibitem{14} A. Carpio, L.L. Bonilla, Edge Dislocations in Crystal Structures Considered as Traveling Waves in Discrete Models, Phys. Rev. Lett. 90 (2003) 135502.

\bibitem{15} S.H. Strogatz, R.E. Mirollo, Phase-Locking and Critical Phenomena in Lattices of Coupled Nonlinear Oscillators with Random Intrinsic Frequencies, Physica D 31 (1988) 143-168.

\bibitem{16} Z. Zheng, B. Hu, G. Hu, Collective phase slips and phase synchronizations in coupled oscillator systems, Phys Rev E 62 (2000) 402-408.

\bibitem{17} H.F. El-Nashar, A.S. Elgazzar, H.A. Cerdeira, Nonlocal synchronization in nearest neighbour coupled oscillators, Int. J. Bifurcation Chaos Appl Sci Eng 12 (2002) 2945-2955.

\bibitem{18} H.F. El-Nashar, Y. Zhang, H.A. Cerdeira, F.A. Ibyinka, Synchronization in a Chain of Nearest Neighbors Coupled Oscillators with Fixed Ends, Chaos 13 (2003) 1216-1225.

\bibitem{19} H.F. El-Nashar, Phase Correlation and Clustering of a Nearest Neighbor Coupled Oscillators System, Int. J. Bifurcation Chaos Appl. Sci. Eng. 13 (2003) 3473-3481.

\bibitem{20} P. Muruganandam, F.F. Fereira, H.F. El-Nashar, H.A. Cerdeira, Analytical calculation of the transition to complete phase synchronization in coupled oscillators, Pramana J. Phys. 70 (2008) 1143-1151.

\bibitem{21} H.F. El-Nashar, P. Muruganandam, F.F. Ferreira, H.A. Cerdeira, Transition to complete synchronization in phase-coupled oscillators with nearest neighbor coupling , Chaos 19 (2009) 013103.

\bibitem{22} H.F. El-Nashar, H.A. Cerdeira, Determination of the critical coupling for oscillators in a ring, Chaos 19 (2009) 033127.

\bibitem{23} D. Botez, D.R. Scifers, Diode Laser Arrays, Cambrige
University Press, New York, 2005.


\bibitem{24} K. Chang, L.-H. Hsieh, Microwave Ring Circuits and Related
Structures, John wiley and Sons Inc., New Jersey, 2004.

\end{thebibliography}
\end{document}